\documentclass[a4paper,english,prl,twocolumn]{revtex4}
\usepackage[T1]{fontenc}
\usepackage[latin9]{inputenc}
\usepackage{graphicx}

\makeatletter
\@ifundefined{textcolor}{}
{%
 \definecolor{BLACK}{gray}{0}
 \definecolor{WHITE}{gray}{1}
 \definecolor{RED}{rgb}{1,0,0}
 \definecolor{GREEN}{rgb}{0,1,0}
 \definecolor{BLUE}{rgb}{0,0,1}
 \definecolor{CYAN}{cmyk}{1,0,0,0}
 \definecolor{MAGENTA}{cmyk}{0,1,0,0}
 \definecolor{YELLOW}{cmyk}{0,0,1,0}
 }

\makeatother

\usepackage{babel}

\begin{document}

\preprint{This line only printed with preprint option}

\title{Comment on "Doping Driven ($\pi,0$) Nesting and Magnetic Properties
of Fe$_{1+x}$Te Superconductors" [Phys. Rev. Lett. 103, 067001 (2009)]}

\author{Prabhakar P. Singh}

\affiliation{Department of Physics, Indian Institute of Technology Bombay, Powai,
Mumbai- 400076, India }


\maketitle
In a recent Letter \citep{MJHan2}, Han and Savrasov described the
changes in the electronic structure and magnetic properties of Fe$_{1+x}$Te
as a function of $x$ using the full-potential, linear muffin-tin
orbital (FP-LMTO) method within the rigid-band approximation (RBA).
In particular, they used the electronic structure of FeTe and RBA
to study the band structure, density of states (DOS), and Fermi surface
(FS) of Fe$_{1.068}$Te (or Fe$_{1.063}$Te) and Fe$_{1.141}$Te.
They find that the excess Fe drives $(\pi,\pi)$ FS nesting in FeTe
to a $(\pi,0)$ nesting in Fe$_{1.068}$Te. 

In this Comment, using Korringa-Kohn-Rostoker coherent-potential approximation
method in the atomic-sphere approximation \citep{PPSingh-fese} (KKR-ASA
CPA) to describe the effects of disorder due to excess Fe in FeTe
alloys, we show that (i) the rigid-band approximation is inadequate
to describe the effects of disorder and its application leads to an
incorrect description of the underlying physics, (ii) the rigid-band
energy shift of $\sim0.76$ eV for going from FeTe to Fe$_{1.068}$Te
(or Fe$_{1.063}$Te), as obtained in Ref. \citep{MJHan2}, is inconsistent
with our FP-LMTO as well as the KKR-ASA results, and thus the FS of
Fe$_{1.063}$Te shown in Fig. 3(b) of Ref. \citep{MJHan2} is not
correct. In addition, Fig. 3 of Ref. \citep{MJHan2} does not show
the FS in $a-b$ plane but it rather gives the top view (slightly
tilted towards top-right) of the full FS, and in the caption, the
positions of the $\Gamma$ and M points have been interchanged. 

\begin{figure}
\includegraphics[clip,scale=0.35]{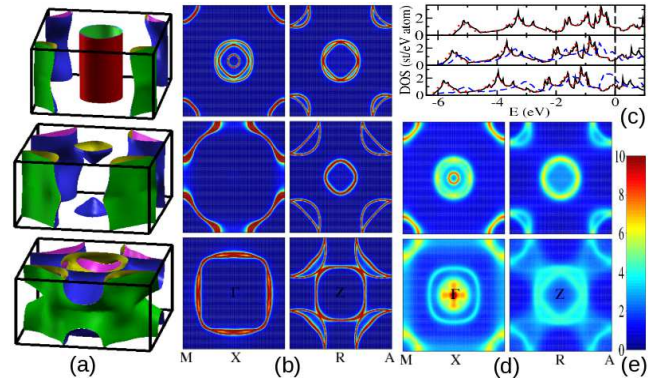}\caption{(Color online) (a) The Fermi surface of FeTe (top), Fe$_{1.068}$Te
(middle) and Fe$_{1.141}$Te (bottom) using FP-LMTO method and the
RBA. In FeTe, there is a small, closed cylinder-like structure at
the center, which is not visible. (b) The FS of FeTe (top), Fe$_{1.068}$Te
(middle) and Fe$_{1.141}$Te (bottom) in $\Gamma$-X-M and Z-R-A planes,
calculated with KKR-ASA in the RBA. (c) The DOS of FeTe (top), Fe$_{1.068}$Te
(middle) and Fe$_{1.141}$Te (bottom) obtained using FP-LMTO (black,
solid) and KKR-ASA (red, dot)  within the RBA. The CPA DOS (blue,
dash) for Fe$_{1.068}$Te and Fe$_{1.141}$Te are also shown. (d)
The FS of Fe$_{1.068}$Te (top) and Fe$_{1.141}$Te (bottom), calculated
with KKR-ASA CPA method. (e) The colormap used in (b) and (d). \label{fig:all}}

\end{figure}

We show in Fig. \ref{fig:all}, the DOS and the FS of Fe$_{1+x}$Te
alloys calculated with the FP-LMTO and the KKR-ASA methods using the
RBA, and with the KKR-ASA CPA method. For $x=0$, our results are
in agreement \citep{epaps} with that of Ref. \citep{MJHan2,ASubedi},
which confirms the reliability of the ASA in the present context.
For going from FeTe to Fe$_{1.068}$Te using the RBA, we need to accommodate
0.544 excess electrons per Fe atom, requiring an upward shift in the
FP-LMTO (KKR-ASA) Fermi energy, $E_{F}$, of $\sim0.39$ $(\sim0.38)$
eV. Similarly, the shift for Fe$_{1.141}$Te is found to be $\sim0.68$
$(0.67)$ eV. Within the RBA, our FP-LMTO and KKR-ASA results are
in good agreement with each other. However, Han and Savrasov find
a shift of $\sim0.76$ eV, almost twice as much as we do, for Fe$_{1.063}$Te.
Note that their shift is close to our shift for Fe$_{1.141}$Te. As
a result the FS of Fe$_{1.063}$Te as shown in Fig. 3(b) of Ref. \citep{MJHan2}
is not consistent with the RBA. Our FP-LMTO and KKR-ASA results for
the FS of Fe$_{1+x}$Te alloys, using the RBA, are shown in Fig. \ref{fig:all}(a)-(b),
and  for $x=0$, they are in agreement with Refs. \citep{MJHan2,ASubedi}.
Not surprisingly, Fig. 3(b) of Ref. \citep{MJHan2} is closer to the
FS of Fe$_{1.141}$Te as shown in our Fig. \ref{fig:all}(a).

We find the KKR-ASA CPA DOS of Fe$_{1.068}$Te and Fe$_{1.141}$Te,
shown in Fig. \ref{fig:all}(c), to be very different from the rigid-band
results of both the FP-LMTO and the KKR-ASA methods. The CPA DOS moves
up and there is a substantial redistribution of states throughout
the energy range, especially around $E_{F}$, making the RBA quite
inaccurate. The inadequacies of the RBA are further revealed by comparing
the CPA FS, shown in Fig. \ref{fig:all}(d) with that of the rigid-band.
The diffused intensity in Fig. \ref{fig:all}(d) indicates the effects
of disordering due to excess Fe in FeTe alloys.

\end{document}